\documentstyle[preprint,12pt,aps,prb]{revtex}
\begin{document}
\newcommand{\mgb}{MgB$_{2}$}
\newcommand{\cm}{$cm^{-1}$}
\newcommand{\tc}{$T_c$}
\title {Evidence for Two Superconducting Gaps in MgB$_2$}
\author{X. K. Chen, M. J. Konstantinovi\'c and J. C. Irwin}
\address{Department of Physics, Simon Fraser University, Burnaby, British Columbia, V5A 1S6, Canada}
\author{D. D. Lawrie and J. P. Franck}
\address{Department of Physics, University of Alberta, Edmonton, Alberta, 
T6G 2J1, Canada}

\maketitle

\begin{abstract} 

We have measured the Raman spectra of polycrystalline {\mgb} from 25 {\cm} 
to 1200 {\cm}. 
When the temperature was decreased below the superconducting transition temperature $T_c$, we observed a 
superconductivity-induced redistribution in the electronic Raman continuum.  
Two pair-breaking peaks appear in the spectra, 
suggesting the presence of two superconducting gaps. 
Furthermore, we have analyzed the measured spectra using a quasi two-dimensional model 
in which two s-wave superconducting gaps open on two sheets of Fermi surface.  
For the gap values we have obtained $\Delta _1 = 22 cm^{-1}$ (2.7 meV) and 
$\Delta _2 = 50 cm^{-1}$ (6.2 meV).  
Our results suggest that a conventional 
phonon-mediated pairing mechanism occurs 
in the planar boron $\sigma$ bands and is responsible for the superconductivity of {\mgb}.

\end{abstract}


\newpage

The recent discovery \cite {a1}  of superconductivity in {\mgb} at temperatures below                 
39K has generated a great deal of interest \cite {a2}.   
The electronic band structure of {\mgb} has been calculated by a number of groups 
who have arrived at a consistent picture for the Fermi surface \cite {a3,a4,a5}.  
It appears that the carriers involved in superconductivity are associated 
with the planar boron $\sigma$ orbitals that give rise to a Fermi surface consisting of two 
approximately cylindrical sheets.  
Carriers in the $\sigma$  bands are predicted to be strongly 
coupled to phonons that involve an appropriate motion of the boron atoms.  
Conventional superconductivity with an s-wave gap, or multiple gaps \cite {a6},
has been predicted, and the unusually high $T_c$ = 40K is 
obtained with intermediate coupling strength \cite {a5}.   
On the other hand, the experimental 
results are quite inconsistent, have yielded conflicting evidence in some cases, and have 
not provided confirmation for the theoretical predictions.  In particular, the magnitude of the 
superconducting gap ($\Delta $) has been measured with techniques including NMR, tunneling, 
high-resolution photoemission spectroscopy, scanning tunneling spectroscopy, IR reflectivity 
and specific heat measurements and the results vary from 2 to 8 meV  \cite {a7,a8,a9,a10,a11,a12,a13}.  
The discrepancies between these results and the lack of knowledge about the k-dependence of the 
superconducting gap have become a major difficulty in understanding the nature of the 
superconducting mechanism of {\mgb}.

Raman spectroscopy is an effective technique for studying the superconducting 
gap in superconductors \cite {a14,a15,a16}.  
Typically, at temperatures below $T_c$, a superconductivity-induced 
renormalization in the Raman spectrum occurs as a result of the opening of the 
superconducting gap.  In particular, the intensity of the electronic Raman continuum is 
depleted at low frequencies because of the lack of available electronic states below the 
gap.   Moreover, scattering from electronic excitations across the gap usually results in 
the appearance of a pair-breaking peak and the peak frequency is a measure of the size of 
the gap ($2\Delta $).  When the gap is anisotropic, its k-dependence can be determined by 
investigating the polarization dependence of the Raman spectra below $T_c$, as different 
scattering geometries essentially probe different portions of the Fermi surface \cite {a16}.  
In fact, 
Raman spectroscopy has been successfully used in the studies of the superconducting gap 
in both conventional \cite {a14} and high-$T_c$ superconductors \cite {a15,a16}. 

In order to determine the magnitude and k-dependence of the superconducting 
gap in {\mgb}, we have carried out an experimental investigation of {\mgb} using Raman 
spectroscopy.  We observed two pair-breaking peaks in the low-temperature spectra, which
indicates the existance of two superconducting gaps.
Moreover, to determine the gap energies and gain insight into the 
superconducting mechanism, we have analyzed the experimental data using a quasi 
two-dimensional model. The results are consistent with expectations based on the 
calculated band structure. In this letter, we report the results of our investigation and 
discuss their physical significance.  

Polycrystalline {\mgb} samples used in this study were prepared by the rapid 
reaction of stoichiometric quantities of powdered amorphous boron (B) and magnesium 
(Mg). The samples show very clean powder x-ray diffraction patterns with all major 
{\mgb} peaks the same as those observed by Nagamatsu $et$ $al$ \cite {a1}.  
The critical temperature 
as determined by resistivity and susceptibility measurements is $T_c = 38.75 K$ and the 
resistive transition width is $\Delta T_c =0.23 K$ ($ 10 - 90 \% $).
The fine structure of the samples was examined with a scanning 
electron microscope and the size of the crystal grains was found to range from 0.15 $\mu m$ to 
0.3 $\mu m$.
 
The Raman spectra were obtained in a quasi backscattering geometry using the 
514.5 nm line of an Ar-ion laser, which was focused onto the sample with a cylindrical 
lens to provide an incident intensity of about $1 W/cm^2$. In particular, the incident laser power was 0.2 mW and the size of the sampling spot was about $ 40 \mu m \times  600 \mu m $. By examining the temperature and laser power dependence of the spectra, the laser heating effect was estimated to be 
$5 \pm 3 K$.  The temperatures reported in this paper are the 
ambient temperatures without any correction. Two polarization configurations were used in the 
Raman measurements, VV and HV, where VV (HV) represents the scattering geometry 
in which the polarization direction of the incident light is vertical (horizontal), and the 
polarization direction of the scattered light is always vertical.  

Fig 1 shows the VV and HV Raman spectra of {\mgb} measured at 15 and 45K. 
All the spectra presented in this paper have been divided by the Bose factor 
(considering the laser heating effect, a 5K correction has been added to the temperature
in the Bose factor). 
The intensity levels of the spectra have been normalized using their integrated intensity in the 
frequency region of interest.  Clearly, there are no well-defined phonon lines in either the 
VV or HV spectra.  Instead, a broad maximum centered at about 620 {\cm} appears in both 
scattering geometries, and changes slightly when the temperature is cooled below $T_c$.  
This broad feature can be interpreted as due to phonon contributions throughout the 
Brillouin zone. The frequency range of this feature is consistent with theoretical 
calculations of the phonon energies \cite {a5,a17} as well as the result of neutron scattering 
measurements of the density of states of phonons \cite {a18}. As we know, in disordered materials phonons away from the Brillouin zone center can become Raman active \cite {a19}.  In this case the phonon density of states is reflected in the Raman spectra. In fact, the broadness of this feature and the similarity between the shapes of the polarized (VV) and depolarized (HV) 
spectra, as well as the intensity ratio of the VV and HV spectra, are very similar to what 
has been observed in amorphous materials \cite {a20}. However, since the X-ray diffraction 
patterns clearly show typical sharp lines of powdered crystals, our samples are unlikely 
amorphous. Moreover, considering the possibility of heterogeneity, we examined 
spectra from different sampling areas but did not find
any significant difference.  
Therefore, we believe that the samples are reasonably homogenous but strongly disordered. 

As shown in Fig 1, when the temperature is decreased below $T_c$, a 
superconductivity-induced redistribution occurs in the low frequency region 
($\omega < 280 cm^{-1}$) of the spectra.  To investigate this redistribution more closely, 
we have subtracted 
the 45K spectra from the 15K spectra and plotted the results in Fig 2.  
Clearly two peaks appear at frequencies of about 50 and 105 {\cm}. 
These peaks are identified as pair-breaking peaks \cite {a14} which originate from 
electronic excitations across the superconducting gap.  
The frequency of a pair-breaking peak is a measure of the 
binding energy of the Cooper pairs, which is two times the gap value (2$\Delta $).  
The presence 
of two pair-breaking peaks in the spectra suggests two superconducting gaps in {\mgb}.  

A careful comparison between the VV and HV spectra in Fig 2 reveals that the 
shape of the spectrum is independent of polarization configuration.  
Such polarization independence can be interpreted as the consequence of several possible 
reasons, such as the disorder of the material 
and the isotropy of the superconducting gap.  In the following data analysis, we will 
focus on the VV spectrum and ignore the polarization dependence of the Raman tensor.

According to Klein and Dierker \cite {a14}, in the small wave vector limit ($q \rightarrow 0 $), 
the photon cross section for Raman scattering from pairs of superconducting quasiparticles at 
zero temperature is given by

\begin{equation}
\frac{d^2R}{d\omega d\Omega }=\frac{4Nr_0^2}{\omega}
<\frac{\mid \gamma({\bf k})\mid ^2 \mid \Delta({\bf k})\mid ^2}
{\sqrt {\omega ^2 - 4\mid \Delta({\bf k})\mid ^2}}>
\end{equation}
where N is the density of states for one spin, $r_0$ is the Thomson radius,  
$\gamma ({\bf k})$  is the polarization-dependent Raman 
vertex, and the brackets denote an average over the Fermi surface (FS) with the restriction
$\omega ^2 > 4\mid\Delta({\bf k})\mid^2$.  
Kortus $et$ $al$ \cite {a3} have calculated the electronic 
band structure and mapped out the FS of {\mgb}.  We are particularly 
interested in the two sheets of cylindrical FS from the planar boron 
$\sigma $ bands near the $\Gamma $ 
point.  For a given $k_z$, since these surfaces are isotropic in the $k_{x}-k_{y} $ plane, 
we simply 
assume that two isotropic s-wave gaps ($\Delta _1$ and $\Delta _2$) open on these two sheets of FS.  Then, 
equation (1) becomes,

\begin{equation}
\frac{d^2R}{d\omega d\Omega }=
\left\{ \begin{array} {ccc}
0  & \quad & \omega < 2\Delta _1 \\
\frac{ S_1 ^2 \Delta _1 ^2}{\omega \sqrt {\omega ^2 - 4 \Delta _1 ^2}} & \quad & 2\Delta _2 > \omega > 2\Delta _1 \\
\frac{ S_{1} ^2 \Delta _{1} ^2}
{\omega \sqrt {\omega ^2 - 4 \Delta _{1} ^2}}
+
\frac{ S_{2} ^2 \Delta _{2} ^2}
{\omega \sqrt {\omega ^2 - 4 \Delta _{2} ^2}} & \quad & \omega > 2\Delta _2 \\
\end {array} \right.
\end{equation}
where $S_1$ ($S_2$) is proportional to the area of the first (second) sheet of the FS.  As shown in 
the inset of Fig 2, the calculated spectrum using equation (2) qualitatively agrees with the 
measured spectra, i.e., two peaks at $2 \Delta _1$ and $2 \Delta _2$ and a vertical slope at 
$2 \Delta _1$.  To achieve a more detailed 
agreement, we would need to take into account the slight dispersion along the $k_z$
direction \cite {a3,a4,a5} and the coupling effect between the two $\sigma $ bands \cite {a6}.  
As an approximation, we simply convolute the calculated spectrum with a 
Gaussian function whose HWHM is 7 {\cm} for contributions from $S_1$ and 3.5 {\cm} for 
those from $S_2$.  As shown in Fig 2, this yields a reasonably good fit with the 
experimental data. The parameters obtained from the best fit to the VV spectrum are: 
$S_1$=125, $S_2$=42, $\Delta _1$=22{\cm}, $\Delta _2$=50{\cm}.  
After multiplying by an intensity scaling factor 
of 0.52, the same parameters also yield a good fit to the HV experimental data (see Fig 
2).  

Ideally, for isotropic s-wave gaps, there should be a complete depletion \cite {a14} in the low temperature 
Raman spectra at frequencies below the smaller gap 
because of the lack of available electronic states.   
As shown in the inset of Fig 2, 
the Raman scattering intensity at frequencies below ($ 2\Delta _1 $) should be zero and 
the spectrum has a vertical slope at $ 2\Delta _1 $ .
However, such a depletion can not be seen in the spectra shown in Fig 1.
In fact, the scattering intensity at low frequencies is not zero in any of the spectra. 
Apparently the low-frequency spectra include significant contributions of scattering from 
excitations other than the superconducting quasiparticles. Such excitations could arise from
normal electrons in the $ \pi $ bands, or could be due to the presence of normal state material
in the samples. 
To remove these unwanted contributions from the spectra, we subtracted the 45K
spectra from the 15K spectra.  The results, more accurately 
representing the scattering spectra from superconducting quasiparticles, are shown in Fig 2.
A nearly vertical slope appears slightly below $ 2\Delta _1 $, which is a feature expected for isotropic s-wave gaps \cite {a14,a16}.

Fig 3 shows the temperature dependence of the electronic Raman spectrum 
measured in HV scattering geometry.  Apparently the two pair-breaking peaks 
are absent at temperatures above Tc. When the temperature is decreased below Tc, the spectra 
gradually gain weight in the low frequency region and the two pair beaking peaks are well developed at 15K. 
Recently Bouquet et al \cite {a21} reported
specific heat data which can be fit to a two-gap model. Also, after the submission of this paper,
F. Giubileo et al \cite {a22} (scanning tunneling) and P. Szabo $et$ $al$ \cite {a23} (point-contact spectroscopy) reported
experimental data showing two superconducting gaps with approximately the same gap values as we have obtained. These data indicate that the two gaps open simultaneously at Tc and their temperature dependence agrees with BCS theory, which is consistent with our results.

In summary, we have measured the Raman spectra of polycrystalline {\mgb}. A 
broad maximum centered at 620 {\cm} arises from phonon scattering throughout the 
Brillouin zone and suggests that the samples are strongly disordered.  When the 
temperature is decreased below $T_c$, a superconductivity-induced redistribution in the 
electronic Raman spectrum occurs as a result of the opening of the superconducting 
gaps.  
The appearance of two pair-breaking peaks indicates the existence of two
superconducting gaps in {\mgb}.
Furthermore, we have analyzed the 
experimental data using a quasi two-dimensional model based on the theoretical 
calculations of Kortus $et$ $al$ \cite {a3}.   
The results suggest that two s-wave gaps open on two 
sheets of the Fermi surface from the planar boron $\sigma $ bands near the $\Gamma $ point with gap 
values of $\Delta _1$=22{\cm} (2.7 meV) and $\Delta _2$=50{\cm} (6.2 meV) respectively, 
i.e., 2$\Delta _1$/$k_B T_c$=1.6 
and 2$\Delta _2$/$k_B T_c$=3.7. 
Note that the values of the two gaps embrace a range covering most
of the results of the reported gap measurements \cite {a7,a8,a9,a10,a11,a12,a13}.         
The result of our data analysis suggests that both gaps are conventional s-wave gaps and a BCS phonon-mediated pairing mechanism occurring in the planar boron $\sigma $ bands is responsible 
for the superconductivity of {\mgb}.

The financial support of the Natural Sciences and Engineering Council of Canada 
is gratefully acknowledged.

\begin{figure}
\caption
{Raman spectra of polycrystalline {\mgb} (after dividing by the Bose factor).}
\label{fig1}
\end{figure} 
\begin{figure}
\caption
{Low temperature electronic Raman continua obtained by subtracting the 45K 
spectra from the 15K spectra.  The thick solid lines are theoretical fits.  The HV spectra have been shifted downward by 20 units. 
The inset is a  
calculated spectrum without convoluting with a Gaussian function.}  
\label{fig2}
\end{figure} 
\begin{figure}
\caption
{HV electronic Raman continua obtained by subtracting the 60K spectrum. The 15K, 25K, 30K, 35K and 45K spectra have been shifted upward by 20, 40, 60 and 80 units respectively}  
\label{fig3}
\end{figure}

\end{document}